# Role of Heterogeneities in Staebler-Wronski Effect

S.C. Agarwal
Department of Physics
Indian Institute of Technology Kanpur 208016

Abstract

The effect of light soaking (LS) on the properties of hydrogenated amorphous silicon presents many challenging puzzles. Some of them are discussed here, along with their present status. In particular the role of the heterogeneities in LS is examined. We find that for the majority of the solved as well unsolved puzzles the long range potential fluctuations arising from the heterogeneities in the films can provide answers which look quite plausible.

## 1. Introduction

Among the amorphous semiconductors, hydrogenated amorphous silicon (a-Si:H) is the only one so far, which can be doped efficiently. This has made it possible to make devices from it, e.g. solar-cells. The properties of a-Si:H have made these solar cells a prime candidate as a renewable source of energy. The cost considerations have not yet allowed them to replace the conventional sources of electricity, and the efficiency of the solar cells needs to be improved to make them cost competitive. The situation becomes worse when one realizes that the efficiency of these solar cells decreases when they are exposed to the sun. This effect is known as the Staebler Wronski Effect (SWE), after its discoverers [1] Staebler and Wronski. They found that a-Si:H degrades upon light soaking (LS) and goes into a metastable state with lower dark and photo conductivities, which can be annealed back at about $200°C$. Because of the interest in a-Si:H as a Solar Cell material, many workers around the world have been studying this phenomena (SWE) in many different ways, trying to eliminate it. However, even after years of research, carried out by more than a generation of workers, not only the degradation persists, its proper understanding is still missing. In several articles from time to time, several authors have looked at the status of our understanding. Amongst these, the reviews by Fritzsche [2-3] are particularly illuminating, and describe quite clearly the various explanations provided to understand the experimental results and the puzzles associated with them. In what follows, we describe some of these puzzles and take a closer look at them with a view to understand the role of heterogeneities present in a-Si:H.

.

Let us first recall our beliefs that are generally accepted as established for SWE.

1. The energy released in the recombination of photogenerated electron hole pairs creates defects and causes degradation.
2. LS results in creation of dangling bonds that are produced probably by breaking Si-H bonds or other weak bonds, and the dangling bonds are the main defects that are important (false!).
3. Movement of hydrogen takes place for the creation of dangling bonds and for blocking their annihilation.

It is now well accepted that even the best quality films of a-Si:H have an inhomogeneous distribution of hydrogen, giving rise to density fluctuations. The NMR experiments of Reimer et al. [4] show the evidence of clustering. Many authors have pointed out the importance of heterogeneities in the creation of metastabilities in a-Si:H upon light soaking known as Staebler Wronski Effect (SWE). We take the inhomogeneities into account by measuring the widths of the Long Range Potential Fluctuations (LRPF) created by these inhomogeneities. We shall show here how the consideration of LRPF can explain several observations, related with light soaking. But before doing that we give a brief introduction to the LRPF, in the next section.

## 2. Heterogeneities and Long Range Potential Fluctuations

Fig.1 shows a two dimensional representation of potential fluctuations that might be present in a [5-6] heterogeneous material, like a-Si:H. The xy plane is the space and the vertical z-axis is the energy.. The potential fluctuations are long range (~10 nm), so as to not to allow tunneling and are the result of charge transfer from one local region to a nearby region, which may have a different chemical potential or different band gap owing to inhomogeneities. It is well established experimentally [7] as well as theoretically [8] that hydrogen increases the band gap in a-Si:H and also that hydrogen is distributed inhomogeneously [4] in a-Si:H. [The increase in band gap in the theoretical calculations with as small a sample as 216 Si atoms and this result should therefore hold for the 10 nm cube, the range of the LRPF] The material may therefore be thought of a collection of heterojunctions in which the distance between the fluctuating bands changes from place to place, as shown in Fig. 1. These are called 'uncorrelated' LRPF [9] and are caused by density fluctuations. There are also the 'correlated' ones, in which both the bands are affected equally, leaving the distance between them unchanged. In Fig.1, $E_C$ and $E_V$ represent the percolation thresholds of electron and holes, respectively. An electron with energy $\geq E_C$ can find a continuous path from the left to right hand side by going around the potential hills.

We would like to point out that Fig. 1 shows both the correlated as well as uncorrelated fluctuations. This is somewhat different than the more familiar picture [6], which shows only the correlated (symmetric) part.

Fig.2 shows how the conduction path may change as a function of energy of the electron. Here 'A' stands for the allowed regions whereas label 'P' means the prohibited (shaded) region. We see that as E increases from ($E<<E_C$) 2(a) to 2(c), first a percolation channel is formed [2(b)] ($E \approx E_C$) and then for $E >> E_C$ the whole sample participates in conduction [2(c)]. The isolated regions marked 'L' are electrons in the localized states as they cannot conduct. [For a detailed discussion of the various consequences of this model please see Ref. 6] Here we would like to point out only a few relevant differences between this 'fluctuating band model' and the usual 'rigid band model'. First, the localized and non-localized states can coexist at the same energy, in the fluctuating band model. Second, in this model the mobility gap and the optical gap may be different. And finally the same states which behave like extended in optical absorption may look localized in transport. Therefore, it is not justified to compare the numbers obtained from the non-transport measurements (e.g. ESR, optical absorption, etc) with those from the transport measurements (conductivity, photoconductivity).

## 2.1 Measuring the width of the LRPF:

It is generally observed that for band conduction the slopes of log σ vs 1/T and S vs 1/T graphs are equal in crystals but not in amorphous semiconductors. Overhof and Beyer [10-11] argued that LRPF are responsible for this difference. In order to eliminate the statistical shift of the Fermi level ($E_F$), they constructed a function Q(T) which is a linear combination of σ and S and is independent of $E_F$. Using a model and computer calculations they showed that under certain assumptions (likely to be valid here), the slope $E_Q$ of Q vs 1/T plot is related to the width Δ of the potential fluctuations in the amorphous semiconductor. For the correlated LRPF, their procedure gave $E_Q = 1.25\Delta$, which provides a way to treat these LRPF quantitatively. Unfortunately, however, a similar relation obtained using the same procedure [10], gives physically unacceptable values for Δ. Although, qualitatively the proportionality of the two quantities can be argued, it prevents us from being too exact. This is a serious problem because in the case of LS we are more concerned with the e-h recombinations at the minima, which are generated only by the uncorrelated LRPFs. The correlated kinds tend to prevent recombinations by keeping the photo generated carriers apart.

### 3. LRPF and light soaking

It is known that the energy released during the recombination of photogenerated electron hole pairs, causes the SWE by creating defects, including dangling bonds. These carriers drift under the influence of LRPF and electrons tend to get to the minima in the conduction band whereas holes try to get to the maximum in the valance band. The recombination occurs preferentially in these regions where the local band edges are nearest to each other energetically as well as spatially (e.g. the transitions marked 3 in Fig.1). It may be recalled that hydrogen is distributed non-uniformly in a-Si:H and therefore the local band gap is different in different regions. Since hydrogen increases the gap [8], the places where the recombinations are more likely, must have a smaller concentration of hydrogen. The energy released by the recombination moves hydrogen out of this already hydrogen deficient region and creates dangling bonds. As LS continues the amount of hydrogen left decreases. This is the reason that SWE gets saturated. Further, this should make the band gap narrower at such points, which means that the width of the LRPF should increase upon LS. This has already been measured [12-15] and found to be true. For instance, Fig. 3 shows [13] that the slope $E_Q$ of Q increases from about zero value in the annealed slow cooled state (indicated by +) to 0.14 eV in the LS state (•) for a Li doped a-Si:H film

Reddy et al [16], have found that the width of the potential fluctuations ($\Delta$) and SWE degradation are correlated as they both decrease with the increasing nanocrystalline fraction in their a-Si:H films. This shows that more heterogeneous films with bigger $\Delta$, degrade more. The available literature seems to be consistent with this view. For example, Fig. 4 shows the effect of LS on a-Si:H solar cells prepared at different deposition rates [17]. Note that the cells made at faster rates degrade more. The same authors also reported that the cells made at higher rates were more heterogeneous, as expected.

### 4. Some puzzles examined in the light of LRPF

Now we take up some observations in LS to see how these ideas can help us understand them.

4.1 <u>HYDROGEN CONCENTRATION</u>

➢ <u>Small concentration of hydrogen near dangling bonds:</u> As per the present understanding, the SWE is linked with the presence of hydrogen in a-Si:H. Therefore one might expect to see a lot of hydrogen near the dangling bonds. The local environment of hydrogen and dangling bonds has been studied by Electron Spin Echo Envelope Modulation (ESEEM) of pulsed ESR [18. 19]. With this technique no hydrogen was found closer than 0.4 nm to dangling bonds. The lack of spatial correlation between dangling bonds and hydrogen is a major difficulty [2] in understanding SWE. In fact the density of hydrogen near the dangling bonds created by LS is actually less than the average in the sample. And this is quite intriguing.

As is clear from the foregoing discussion, this difficulty does not arise in the scheme proposed here, because the recombinations take place predominantly at the local minima where there is very little hydrogen to start with. Hence, one expects relatively less hydrogen in the region of dangling bonds/recombination, in agreement with the observations.

.In addition to creating defects like dangling bonds, light can also interact with the silicon lattice and change the bonding arrangements. This can change LRPF and the percolation path.

## 4.2 SEVERAL KINDS OF DEFECTS

Two films may have equal number of defects, but will behave differently, if the defects are of different types. Ample evidence [2] already exists for several kinds of defects. In LRPF model it follows naturally. First there are those defects that are created preferentially at the local LRPF minima during LS. Then there are the native defects which are formed during preparation and can be created almost anywhere. These can be quite different. Further, it may be that light also creates defects other than dangling bonds which affect mainly the transport properties by changing the percolation path.

- ➤ Han and Fritzsche [20] did an experiment to show that $\mu\tau$ and N are not related in a unique manner [Fig. 5]. A* is the annealed state. Upon LS at 160 K the sample follows the dashed curve to B2 and then to B1 after LS at 300K. At B1 the film is annealed stepwise at 430K and now it follows a different path B1 to A. A* is reached after annealing at 480K.

This can be explained in our scheme, by noting that $\mu\tau$ is measured by transport whereas N is measured by optical absorption. Therefore they need not be related, as pointed out earlier. Further, they may be controlled by different sets of defects as envisaged by LRPF. So N is of at least two kinds, in agreement with the conclusion of the authors.

- ➤ Photoconductivity decreases fast initially, then slowly.

Answer may be that there are defects other than the dangling bonds that are formed by the changes in the whole Si structure itself, like strains, stress etc that might reduce the conductivity. Creation of dangling bonds is a slow and inefficient process. The changes in Si network might change the percolation path (that changes the dark and photoconductivity) which may be faster than the creation of dangling bonds.

- ➤ The light induced annealing [21] (LIA) behavior of two samples having equal number of defects but one having light induced defects (LID) and the other having native defects is found to be quite different. The one having LID anneals out faster and leads to a sample with less no. of defects.(Fig. 6)

This difference is expected since light induced dangling bonds are at locations which are different than those of deposition related dangling bonds.

- ➤ Hauschildt et al. [10] did an elegant experiment way back in 1982, which shows that the effect of light soaking is not to just shift the Fermi level towards the middle of the gap. They annealed a light soaked sample progressively, at 80 degrees C and higher, measuring $E_\sigma$ and $E_Q$ after each anneal. They found that both $E_\sigma$ and $E_Q$ decreased, with annealing. Then they took another sample and changed its $E_\sigma$ by doping with increasing amount of Phosphorous. The $E_Q$ vs. $E_\sigma$ curve was qualitatively different, for the two samples. They argued that since $E_Q$ is independent of $E_F$, the effect of LS cannot be a mere shift of $E_F$, but must also involve a change in the percolation path. In other words light induced defects (LID) are different than the defects created by P doping.

### 4.3 DEFECT CREATION EFFICIENCY LOW

- SWE creates only about $N=10^{17}$ cm$^{-3}$ defects(too few defects with poor efficiency)
  It is generally believed that the SWE creates defects by breaking Si-Si or Si-H bonds. However, the number N of light induced defects (LID) is rather small, considering that there are about $5 \times 10^{21}$ Si-H and $10^{23}$ Si-Si bonds per cm$^3$. Moreover, the creation efficiency of the dandling bonds is quite low. It takes about $10^7$ photons to create one dangling bond [2].

The LRPF model can explain this readily. Here the recombination takes place only at select places, namely where the fluctuating valance band has a maximum and the conduction band a minimum, resulting in a local minimum [for example, transitions marked 3 in Fig 1].So, although the whole sample is exposed to light, only a small fraction of the photons create photo induced defects, resulting in a seemingly low efficiency. The total number of defects generated is limited by the number of recombination centers, because we think that the center is no more able to produce a metastable dangling bond after it is depleted of hydrogen, by recombination. Assuming [10] that the sample consists of cubes of length 10 nm (i.e., the range of potential fluctuations so that the band edges do not fluctuate inside the cube) and that there is a junction at each of the interfaces between the cubes a back of the envelope calculation shows that there will be about $7 \times 10^{17}$ minima per cm$^3$; which is closer to the observed number of defects generated by light soaking.

### 4.4 MORE RESULTS EXPLAINED

- It looks as though the recombination takes place at the internal surfaces of the voids. This is because the most stable solar cells are prepared in such a way that they have the least voids. If so, why does the diffusion constant of bulk fit the data?

This can be easily understood because the improvement in the stability is not because the void surface is less but because the heterogeneities that give rise to LRPF are fewer. Therefore, the gap minima, where the recombinations take place are fewer. These minima need not be on the surface.

Hence, the relevant value of the diffusion constant is the bulk value.

- Current injection produces defects. These defects can be annealed slowly by low current annealing but it does not work for LIDs.

This is understandable if one realizes that the light exposure is seen by the entire sample, whereas the current injection only affects only the areas of the percolation channel. Thus the low current does not anneal the LID as it does not see the entire sample.

## 4.6 Further OBSERVATIONS / **REMARKS**

- Creation efficiency has $t^{1/3} G^{2/3}$ dependence as before:

    We have, for bimolecular recombination

    $$n \propto G/N \text{ and } p \propto G/N$$

The rate of creation of defects (N) is given by:

$$dN/dt \propto np \propto (G/N).(G/N),$$

$$N^2 \, dN \propto G^2 \, dt$$

Therefore, $\quad N \propto t^{1/3} G^{2/3}$, as observed experimentally.

The electron hole pairs do not have to cross any energy barrier to recombine and hence the creation of SWE defects (N) should be independent of temperature, as observed.

- ➢ <u>Number of defects mesured may depend on the method of measurement</u>

Two methods are commonly used:

1. The concentration of dangling bonds can be measured by ESR. Here the surface as well as bulk defects are counted. However the other defects that may be present but are ESR inactive are ignored.
2. Sub gap optical absorption can give the density of defects. This has the advantage that it is sensitive to both ESR active as well as ESR inactive defects. However, the surface states may or may not be included depending on the method of measurement. Whereas photo thermal deflection spectroscopy (PDS) measures absorption from the bulk as well as surface, constant photocurrent method (CPM ignoresthe surface).

Although these explanations are valid, there is another point of view possible as far as LRPF are concerned. In CPM only those transitions are considered as absorptions where the electrons are excited above $E_C$ (transitions marked 1 in Fig.1), and the others like 2 (Fig1) are ignored. PDS, on the other hand counts both as absorption as it measures the heat produced, which happens in both cases. Hence PDS has higher absorption than CPM [See Fig. 7]. Therefore we may find that Draft- The number of defects measured by optical absorption by CPM and PDS may not agree with each other.
- The number of defects measured by optical methods may not agree with those measured by transport.
- Not all LRPF give SWE.  Only uncorrelated types LRPF contribute**.**

## 5. Conclusions

We have examined the role of heterogeneities in the light induced degradation of a-Si:H.  We have taken these heterogeneities into account by considering the long range potential fluctuations (LRPF) generated by them (heterogeneities).

We have used a couple of generally accepted ideas, which look quite reasonable to us.  These are

(i) Hydrogen increases the band gap of a-Si and
(ii) The photo-carriers recombine preferentially at the local minima.

These considerations have not only provided alternative explanations of most experimental observations in terms of LRPF, but also enabled us to understand, some of the so far unresolved mysteries.  We give a few examples.

One of the most important puzzles is the fact that although hydrogen is responsible for the creation of dangling bonds, the density of hydrogen] near the dangling bonds is smaller than the average [2?  Using LRPF and the ideas above, it is easy to see that this is exactly what we should expect, since the local minima are in hydrogen deficient region, and that is where the recombinations take place.

Another unexplained phenomenon considered is the low efficiency of creation of light induced defects.  This seems reasonable, if we realize that the recombinations are likely only near the local minima, because in other places the local field will tend to pull photo carriers apart.

Yet another example of an unresolved problem that was taken up is this.  The data shows more degradation in the films containing more voids.  Hence, one may be tempted to conclude that the recombinations occur at the internal surfaces of the voids.  But then why does the data fit the bulk value of the diffusion constant of hydrogen.  In the present scheme we see that the film that has more voids is more heterogeneous and therefore degrades more and the recombinations are in the bulk of the film at the LRPF minima.

The presence of more than one kind of defects is also a natural consequence of the fluctuating band model, as here the optical absorption and electrical conduction involve different states and can have defects that are different.

In addition, the heterogeneities point of view has been applied to some other observations, for which explanations already exist and is found to work well in most cases.  However, most of this exercise has been qualitative.  The reason is that although we can measure the 'correlated' LRPF by analyzing the transport data, a similar analysis does not work for the 'uncorrelated' LRPF.  This is a big hurdle since they(uncorrelated LRPF) are the ones that produce variations in the band gap that promote recombinations and are mostly responsible for degradation.  The correlated LRPF actually discourage recombinations by pulling the photo generated carriers apart.

So there is a need to work out the theory for handling the uncorrelated LRPF, so that quantitative predictions can be made and tested. Only then can the validity of our ideas be confirmed.

## ACKNOWLEDGEMENTS

> **References**

Figure captions.

Fig.1   LRPF as might be present in a-Si:H. The potential hills and valleys control the motion of of charge carriers which conduct by percolation.  $E_C$ and $E_V$ denote the percolation thresholds for the electrons and holes respectively.  Transitions 1 and 2 denote optical absorptions, while those marked 3 represent recombinations at the local minima (after ref 5).

Fig.2   Allowed (A) and prohibited (P, shaded) regions as might be seen by an electron having energies $E<<E_C$ (a),  $E \approx E_C$  (b) and $E >> E_C$ (c).  L denotes localized states.  As E increases A and L increase and P decreases. At the percolation threshold (fig. b),  $E \approx E_C$, a conducting channel can be seen (after ref 6).

Fig. 3   Q as a function of 1/T for a lithium doped a-Si:H sample, in annealed slow cooled state (+), after LS (•) and after fast thermal quenching FQ (∘). The slope $E_Q$ changes from 0 to 0.14 eV after LS but remains unchanged upon FQ (after ref 13).

Fig. 4   Degradation upon LS of the efficiency of a-Si:H solar cells, prepared at different deposition rates as indicated.  The cells made at higher deposition rates degrade more (after ref. 17).

Fig. 5   A plot of µτ vs optical absorption measured in an a-Si:H film. They do not follow the same path while light soaking (A*→ B2 →  B1) or annealing (B1  → A  →  A*) (after ref 20).

Fig.6   Light induced annealing (LIA) in two samples having the same defect density.  One has native defects and the other has LIDs. LIA works better for the sample with LID.(after ref. 21)

Fig. 7  Optical absorption measured in undoped a-Si:H by PDS and CPM.  The sub gap absorption is higher for PDS (after ref. 22).

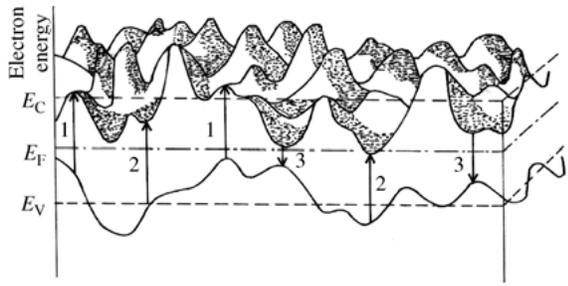

FIG. 1

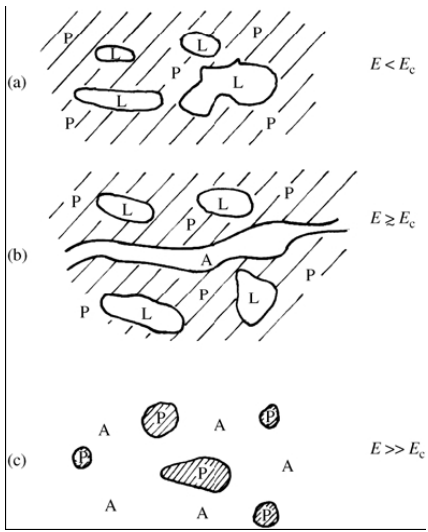

FIG.2

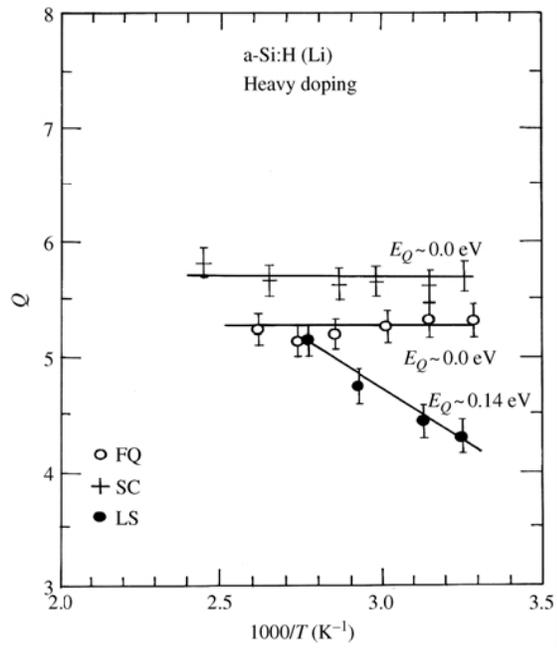

FIG. 3

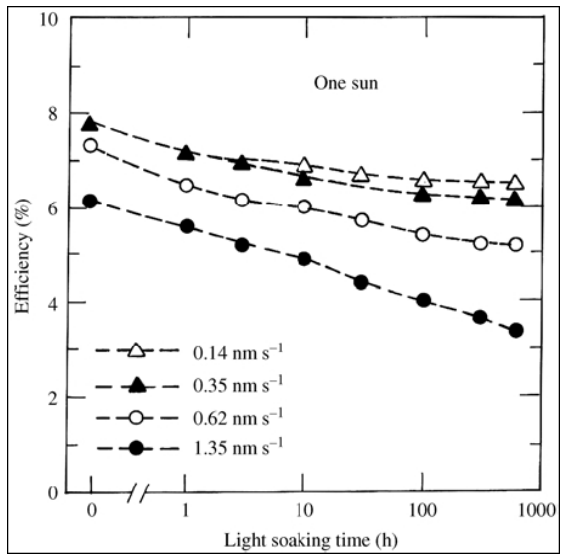

FIG.4

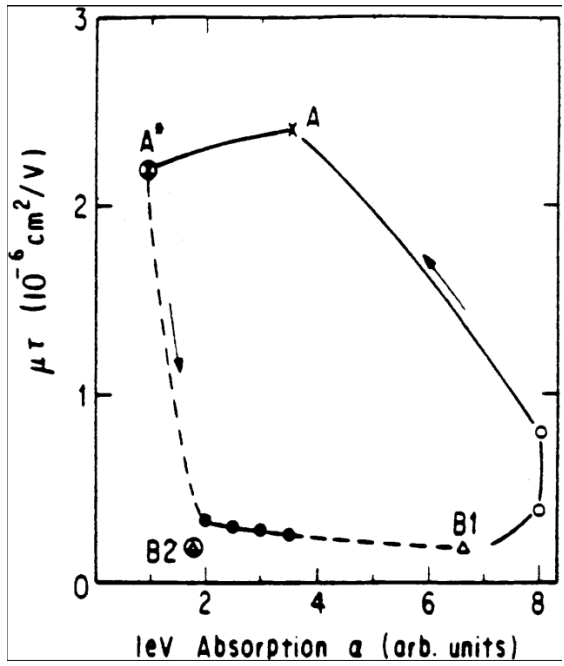

Fig.5

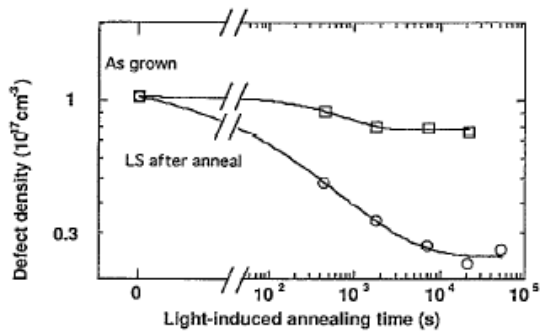

Fig. 6

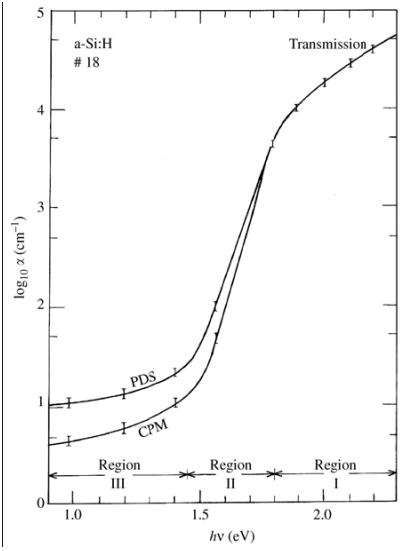

FIG.7